\documentstyle[twocolumn,eqsecnum,aps]{revtex}
\newcommand{\beq}{\begin{equation}}
\newcommand{\eeq}{\end{equation}}
\newcommand{\beqa}{\begin{eqnarray}}
\newcommand{\eeqa}{\end{eqnarray}}
\newcommand{\kB}{\mbox{$k_{\rm B}$}}
\newcommand{\kBT}{\mbox{$k_{\rm B}T$}}
\newcommand{\TR}{\mbox{$T_{\rm R}$}}

\begin{document}
\draft
\title{Vicinal Surface with Langmuir Adsorption: A Decorated Restricted Solid-on-solid  Model  
}
\author{Noriko Akutsu}
\address{Faculty of Engineering, Osaka Electro-Communication
 University,
Hatsu-cho, Neyagawa, Osaka 572-8530, Japan}
\author{Yasuhiro Akutsu}
\address{
Department of Physics, Graduate School of Science, Osaka University,
Machikaneyama-cho, Toyonaka, Osaka 560-0011, Japan}
\author{Takao Yamamoto}
\address{
Department of Physics, Faculty of Engineering, Gunma University,
Kiryu, Gunma 376-0052, Japan}
\date{\today}
\maketitle
\begin{abstract}
We study the vicinal surface of the restricted solid-on-solid model coupled with the Langmuir adsorbates which we regard as two-dimensional lattice gas without lateral interaction. The effect of the vapor pressure of the adsorbates in the environmental phase is taken into consideration through the chemical potential.  We calculate the surface free energy $f$, the adsorption coverage $\Theta$, the step tension $\gamma$, and the step stiffness $\tilde{\gamma}$ by the transfer matrix method combined with the density-matrix algorithm.  Detailed step-density-dependence of $f$ and $\Theta$ is obtained.  We draw the roughening transition curve in the plane of the temperature and the chemical potential of adsorbates.   We find the multi-reentrant roughening transition accompanying the inverse roughening phenomena.  We also find quasi-reentrant behavior in the step tension.

\end{abstract}
%
%

%

\narrowtext

\section{Introduction}

The adsorption effect has been an important subject of study in wide area of surface science\cite{desjonqueres,pimpinelli,copel,eaglesham,chai}.  As for the steps on the surface, either with or without adsorption, detailed experiments in various length scales\cite{ozcomert,fujita,hannon,jones,latyshev,williams98,vonhoegen98} has become possible due to the development of the surface observation technique such as STM (scanning-tunneling microscopy)\cite{stm},  LEEM (low-energy electron microscopy)\cite{leem} and REM (reflection electron microscopy)\cite{rem}. The length scale of STM is the order of nanometer, while that of LEEM and REM is more than hundreds nanometer.  Then, the importance of the detailed statistical mechanical study connecting observations in different length scales has been increasing.

As for the statistical mechanical study of vicinal surface with adsorption, however, reliable calculations have been difficult due to the criticality of the vicinal surface.  In the present paper, to overcome the difficulty, we use the product wave-function renormalization group (PWFRG) algorithm\cite{pwfrg,HOA,OHA} which is a variant of the density matrix renormalization group (DMRG) algorithm\cite{white}.  The efficiency of the method for the vicinal surface problem has been demonstrated in Ref. \cite{pwfrg,HOA,akutsu98}

We make statistical mechanical calculations on the vicinal surface of the restricted solid-on-solid model coupled with the Langmuir adsorbates which we regard as two-dimensional lattice gas without lateral interaction.  We aim at clarifying the ``entropic effect'' for surfaces and steps caused by adsorbates.  The results are applicable to realistic vicinal surfaces if the temperature is high enough to neglect the interaction between the adsorbates, or the interaction itself is negligibly small.

This paper is organized as follows.
In \S II, we introduce a ``decorated'' RSOS model where the ledge formation energy is locally modified by the presence of the adsorbate atom on the ledge.  After tracing out the lattice-gas degrees of freedom, we obtain the effective RSOS model which we show in \S III.  In \S IV, we give phase diagram associated with the roughening transition. We show that the multi-reentrant roughening transition occurs.  Calculated results for surface free energy per projected area $f$, surface gradient $p$, magnetization (adsorption coverage $\Theta$) $M$, step tension $\gamma$, and step stiffness $\tilde{\gamma}$ are given in \S V.  Quasi-reentrant behavior of step tension is also shown.  In \S VI, we give summary of the paper.

\section{Model hamiltonian} 
We consider the square lattice where the position of each site $i$ is expressed in terms of a pair of integers as $i=(m,n)$.  To each site $i$ we assign the integer height variable $h_{i}$ ($=h(m,n)$).  An Ising spin variable $\sigma_{x}(m,n)$ or $\sigma_{y}(m,n)$ representing the adsorbate atom is assigned to each lattice edge ($\sigma_{x}$ for vertical edge, and $\sigma_{y}$ for horizontal edge).  The Hamiltonian of the decorated RSOS model is written as follows:
\begin{eqnarray}
{\cal H}&=&\sum_{m,n}\{ \epsilon [1-\alpha \sigma_y{(m,n)}] \cdot |h{(m+1,n)}-h{(m,n)}|  \nonumber \\
&& + \epsilon [1-\alpha \sigma_x{(m,n)}] \cdot |h{(m,n+1)}-h{(m,n)}|\} \nonumber \\
&&-\eta\sum_{m,n}[h{(m+1,n)}-h{(m,n)}] \nonumber \\
&&  - H \sum_{m,n} [\sigma_x{(m,n)}+ \sigma_y{(m,n)}]\\
&=& {\cal H}_{\rm RSOS}+{\cal H}_{\rm L}+{\cal H}_{\rm int},
\label{hamil}
\end{eqnarray}
where
\begin{eqnarray}
{\cal H}_{\rm RSOS} &=& \sum_{<i,j>} \epsilon |h_i-h_j| \nonumber \\
&&- \eta \sum_{m,n}[h{(m+1,n)}  -h{(m,n)}] ,  \nonumber \\
{\cal H}_{\rm L} &=&  - H \sum_{m,n} [\sigma_x{(m,n)}+ \sigma_y{(m,n)}] ,  \nonumber \\
{\cal H}_{\rm int} &=& - \sum_{m,n} \epsilon \alpha [ \sigma_y(m,n) \cdot |h(m+1,n)-h(m,n)|  \nonumber \\
&& + \sigma_x(m,n) \cdot |h(m,n+1) - h(m,n)| ].
\label{hamil3}
\end{eqnarray}
The Hamiltonian ${\cal H}_{\rm RSOS}$ describing the surface configuration of the mother crystal is that of the restricted solid-on-solid (RSOS) model\cite{RSOS} on the square lattice with nearest-neighbor (nn) interactions; $\epsilon$ is the ``bare'' ledge energy, and $\eta$ the Andreev field\cite{andreev} to make surface tilt\cite{akutsu98,akutsu99}.  In (\ref{hamil}) and (\ref{hamil3}), the height difference $h_i - h_j$ ($\equiv \Delta h_{ij}$) is restricted to be $\Delta h_{ij}=0,\pm1$ (RSOS condition), which is a reasonable simplification because configurations with large $|\Delta h_{ij}|$ are energetically unfavorable. 

 The Hamiltonian ${\cal H}_{\rm L} $ describes a system of free Ising spins ($\sigma = \pm 1$) on the edges of the square lattice (Fig. \ref{vertex}).  We regard these free spins as the Langmuir adsorbates in a monolayer adsorption without lateral interaction. The adsorption energy is represented by $H$.  Since the adsorbates are situated at the ``bridge'' sites, the total number of adsorption sites are twice as many as the total number of the lattice points of the RSOS model. 

The Hamiltonian ${\cal H}_{\rm int}$ represents the interaction between the RSOS system and the adsorbate, with $\alpha$ being a parameter describing the local change of ledge energy caused by a spin (an adsorbate atom)~\cite{ladeveze}.  In what follows, we assume $\alpha>-1$.

The decorated RSOS model is a special case ($J=0$) of the RSOS-Ising coupled model\cite{akutsu99,akutsu00} where the interaction between adsorbates is taken into account.

Since the lattice-gas representation is physically more natural for description of the adsorbates than the spin representation, we map the decorated RSOS to the equivalent RSOS-lattice-gas system.  We introduce the lattice gas (LG) variable $c= \{0, 1\}$ by $c=(\sigma+1)/2$; $\sigma=1$ corresponds to the presence of an adsorbate ($c=1$), and $\sigma=-1$ to the absence of the adsorbate ($c=0$). Then ${\cal H}$ (eq. (\ref{hamil})) is rewritten in terms of the LG variable as follows:
\beqa
{\cal H}_{\rm LG}&=&\sum_{<i,j>} \tilde{\epsilon} |h_i-h_j| -\eta\sum_{m,n}[h{(m+1,n)}-h{(m,n)}] \nonumber \\
&& - \mu \sum_{m,n} [c_x{(m,n)}+ c_y{(m,n)}] \nonumber \\
&&- \sum_{m,n} \tilde{\epsilon} \tilde{ \alpha } [ c_y(m,n) \cdot |h(m+1,n) - h(m,n)|  \nonumber \\
&& + c_x(m,n) \cdot |h(m,n+1) - h(m,n)| ]  \nonumber \\
&& + {\rm const}.
, \label{hamilLG}
\eeqa
where
\beqa
\tilde{\epsilon}&=&\epsilon (1+\alpha),\   \tilde{\alpha}=\frac{2\alpha}{1+\alpha},\       \mu = 2H,  \label{epLG} \\
&&  |h_i-h_j| =\{0,1\}.
\eeqa
The parameter $\tilde{\epsilon}$ represents the microscopic ledge energy of the ``clean'' (i. e., without adsorbate) surface, $\mu$ the surface chemical potential, and $\tilde{\alpha}$ the redefined coupling parameter.  From (\ref{hamilLG}) we see that, for $\tilde{\alpha}>0$ (resp. $\tilde{\alpha}<0$), the adsorbate makes the microscopic ledge energy smaller (resp. larger) than that of the clean surface, which will lead to rougher (resp. smoother) surface with larger (resp. smaller) population of adsorbates on the ledges (Fig. \ref{vertex}).  

Physically, the chemical potential $\mu$ consists of two terms,
\beq
\mu= \epsilon_{\rm ads}+ \kBT \ln P,
\eeq
where $\epsilon_{\rm ads}$ is the adsorption energy which depends on the combination of the atomic species of the adsorbate/substrate, and $P$ is the environmental vapor pressure of adsorbate gas in thermal equilibrium with the surface ($\kB$: Boltzmann constant, $T$: temperature).  Introducing $P_0$ by $\epsilon_{\rm ads}=  -\kBT \ln P_0$, we have
\beq
\mu =  \kBT \ln (P/P_0) =2H.    \label{mu} \\
\eeq

We should note the following symmetry of the system: the Hamiltonian (\ref{hamil}) is invariant under the change $(\alpha,  \sigma_x, \sigma_y, H) \leftrightarrow (- \alpha, -\sigma_x, -\sigma_y, -H)$.  Although the lattice gas representation is more natural than the Ising-spin representation, the latter has a merit that this symmetry is apparently seen.  In the following sections, we adopt the Ising-spin representation.  Results in the lattice-gas representation are also shown if necessary.

\section{Effective RSOS hamiltonian}

Since each nn height difference $\Delta h$ is restricted to $\{0, \pm1\}$, we have an identity $F(|\Delta h|)=F(0)\delta_{\Delta h,0}+F(1)(\delta_{\Delta h,1}+\delta_{\Delta h,-1})$ with $\delta_{i,j}$ being the Kronecker's delta.  Using this identity, we can exactly decimate spin variables $\{\sigma\}$ in ${\cal H}$ to obtain the effective Hamiltonian ${\cal H}^{\rm eff}$ as follows ($\beta=1/(\kBT)$):
\beqa
\sum_{\{ \sigma{(m,n)} \}} &&  \exp(-\beta {\cal H}) \nonumber \\
&=& (2 \cosh \beta H)^{Nz/2}\exp(- \beta {\cal H}^{\rm eff}) , \label{hamileff} \\
\beta {\cal H}^{\rm eff}&=&\sum_{<i,j>} \beta \epsilon^{\rm eff}  |h_i-h_j| \nonumber \\
&& -\eta\sum_{m,n}[h{(m+1,n)}-h{(m,n)}],
\label{heff}\\
\epsilon^{\rm eff} &=& \epsilon - \kBT \ln \left [ \frac{\cosh ( \beta \alpha \epsilon + \beta H)}{ \cosh \beta H} \right ], \label{eeff}
\eeqa
where $N$ is the number of lattice point in the RSOS system and $z$ ($= 4$) is the coordination number.  Hence the decorated RSOS system is equivalent to a non-decorated RSOS system with the effective ledge energy  $\epsilon^{\rm eff}$. 

In Fig. \ref{eeff_fig}, we show temperature and chemical potential dependence of effective ledge energy $\epsilon^{\rm eff}$.  We should note that, at sufficiently low temperatures, $\epsilon^{\rm eff}$ is well approximated by
\beq
\epsilon^{\rm eff}= \epsilon (1-|\alpha  + H/\epsilon| + |H/\epsilon|). 
\eeq

We rewrite the $\epsilon^{\rm eff}$ by using $\tilde{\epsilon}$ and $\tilde{\alpha}$ in (\ref{epLG}) as follows:
\beq
\epsilon^{\rm eff} = \tilde{\epsilon}(1-\tilde{\alpha}/2) - \kBT \ln \left [ \frac{\cosh ( \beta \tilde{\alpha} \tilde{\epsilon}/2 + \beta H)}{ \cosh \beta H} \right ]. \label{eefflg}
\eeq
At low temperatures, we have
\beq
\epsilon^{\rm eff}=\tilde{\epsilon} (1-\tilde{\alpha}/2-| \tilde{\alpha} /2 + H/\tilde{\epsilon}| + |H/\tilde{\epsilon}|).\label{low_t_e_eff}
\eeq
In Fig. \ref{eefflg_fig}, we show temperature and chemical potential dependence of $\epsilon^{\rm eff}$ in the lattice-gas representation.

At low temperatures where the entropy effect of adsorbate configurations almost vanishes,  $\epsilon^{\rm eff}$ should become approximately $\tilde{\epsilon} (1-\tilde{\alpha} \Theta )$.  Therefore, from eq. (\ref{low_t_e_eff}), we have the low-temperature expression of $\Theta$:
\beq
\Theta = \frac{1}{2}+ \frac{1}{\tilde{\alpha}} (| \tilde{\alpha} /2 + H/\tilde{\epsilon}| - |H/\tilde{\epsilon}|).
\eeq

\section{Multi-reentrant roughening transition}
\subsection{Determination of roughening transition temperature}

The roughening transition temperature of the original (non-decorated) RSOS model is known to be given by\cite{dennij} $\epsilon/(\kB T_{\rm R}^{\rm RSOS}) = 0.633\ldots \equiv \zeta_{\rm R}$.  Therefore, $\TR$ of the decorated RSOS model (the surface with Langmuir adsorption) is determined from $\epsilon^{\rm eff}/(\kB\TR)=\zeta_{\rm R}$; when $\epsilon^{\rm eff}/\kBT>\zeta_{\rm R}$ (resp. $\leq \zeta_{\rm R}$), the surface is smooth (resp. rough).  From Eq. (\ref{eeff}), we obtain an equation,
\beq
\zeta_{\rm R} = \frac{\epsilon}{\kB\TR} - \ln \left \{ \frac{\cosh [( \alpha \epsilon + H)/\kB\TR]}{ \cosh (H/\kB\TR)} \right \} \label{j0TR},
\eeq
which determines $\TR$ (see Fig.\ref{j0TR_fig}).  For instance, at $\alpha=0.5$ and $H=0$, we obtain $\kB\TR/\epsilon = 1.446$.  From Eq. (\ref{j0TR}) we have
\beq
\frac{H}{\kB\TR} = \frac{1}{2} \ln \frac{\exp[\zeta_{\rm R}-(1+\alpha)\epsilon/(\kB\TR)]-1}{1-\exp[\zeta_{\rm R}-(1-\alpha)\epsilon/(\kB\TR)]}, \label{phaseTR}
\eeq
which allows us to draw the critical line in the $H-T$ plane, as shown in In Fig. \ref{phaseTR_fig}.

In $H\rightarrow\infty$ limit, we have $\langle\sigma\rangle$ (magnetization) $\rightarrow=1$; all of the ledge sites are covered by adsorbates ($\Theta=1$) (Fig. \ref{j0TR_fig} (a)).  In this limit, $\epsilon^{\rm eff}$ becomes $\epsilon (1-\alpha)$ from which the roughening transition temperature becomes $\kB\TR/\epsilon = (1-\alpha)/\zeta_{\rm R}$.  In $H\rightarrow -\infty$ limit, we have $\langle\sigma\rangle\rightarrow=-1$, which corresponds to the clean surface ($\Theta=0$).  In this limit, $\epsilon^{\rm eff}$ converges to $\epsilon (1+\alpha) = \tilde{\epsilon}$, and the roughening transition temperature becomes $\kB\TR/\epsilon = (1+\alpha)/\zeta_{\rm R}$.

\subsection{Multi-reentrant transition}

Around $\alpha=1$, there occurs reentrant phase transition in a range of $H$ (Fig. \ref{phaseTR_fig} (b) and (c)).  At $\alpha=1$, as we decrease temperature fixing $H/\epsilon = 0.01$, the system cause phase transition twice: rough $\rightarrow $ smooth $\rightarrow $ rough (Fig. \ref{j0TR_fig} (b) and \ref{phaseTR_fig} (b)).  This reentrant transition is similar to what has been known in the 2D frustrated Ising models\cite{chikyu,akutsu96}.

At $\alpha=1.1$ and $H/\epsilon = -0.0735$, the phase transition occurs three times as we decrease temperature, which can be seen from Fig. \ref{j0TR_fig} (c) and \ref{phaseTR_fig} (c); the system shows ``multi-reentrant'' roughening transition, which has not been known previously.  The origin of this phenomenon is the non-trivial behavior of $\epsilon^{\rm eff}$ in the parameter region around $\alpha=1$ and $H=0$.  In this region, $\epsilon^{\rm eff}$ varies sensitively with $H$, making the critical curve non-monotonic.

It should be noted that the (multi-) reentrant behavior accompanies the inverse roughening~\cite{luijten94} proposed for CsCl type crystal surface.  Hence, our decorated RSOS model is a different but physically realizable system showing the inverse roughening.

\section{Behavior of the vicinal surface}

\subsection{Universal behavior of the vicinal surfaces}

The exact mapping to the ordinary RSOS system given by (\ref{hamileff})$\sim$(\ref{eeff}) implies that the vicinal surface of the decorated RSOS model belongs to  the Gruber-Mullins-Pokrovsky-Talapov (GMPT) universality class\cite{gmpt,haldane,izuyama,jayaprakash,schults,nolden} below the roughening temperature.  This means that, in spite of the presence of the adsorbates, the coarse-grained vicinal surface is well described in terms of ``effective terraces'', ``effective steps'' and ``effective kinks'' allowing us to take the terrace-step-kink (TSK) picture. 

The surface free energy (per projected area) $f=f(\rho)$ ($\rho$: step density) has the expansion characteristic to the GMPT universality class,
\begin{equation}
f(\rho)=f(0)+\gamma \rho +B \rho^3 +O(\rho^4), \label{GMPTform}
\end{equation}
where $\gamma$ is the step tension and $B$ is the  step interaction coefficient.  When the system is not isotropic, $\gamma$ and $B$ depend on the mean running direction of steps.  By $\theta$, we denote the angle between one of the crystal axes on the facet plane and the mean running direction of steps.\cite{Theta}  We should then write $\gamma = \gamma(\theta)$ and $B = B(\theta)$.  There exists the following universal relation between $\gamma(\theta)$ and $B(\theta)$\cite{aay88,yamamoto88} (which leads to the universal Gaussian curvature jump at the facet edge\cite{aay88,yamamoto88}) for systems where the step-step interaction is short-ranged:
\begin{equation}
B(\theta)=\frac{\pi^2 }{6} \frac{(\kBT)^2}{\tilde{\gamma}(\theta)},\label{univrel}
\end{equation}
where $\tilde{\gamma}(\theta)$ defined by
\begin{equation}
\tilde{\gamma}(\theta)=\gamma(\theta)+\partial^2 \gamma(\theta)/\partial \theta^2
\end{equation}
is the step stiffness.

\subsection{Calculation of surface and step quantities by PWFRG method}

From the Hamiltonian (\ref{hamil}) and the partition function $Z$ associated with it, we obtain the Andreev surface free energy  $\tilde{f}(\vec{\eta})$ as
\beq
\tilde{f}(\vec{\eta})= - \frac{1}{N} \kBT \ln Z, \label{tildef}
\eeq
where $\vec{\eta} = (\eta_x, \eta_y)$ is the Andreev field.  As was shown by Andreev~\cite{andreev}, $\tilde{f}(\vec{\eta})$ directly gives the equilibrium crystal shape (ECS).  In the present study, we set $\eta_x = \eta$ and $\eta_y = 0$; $\tilde{f}(\eta)- \eta$ curve can be regarded as the profile of ECS near (001) facet along the $x$-direction, and the facet size corresponds to the step tension $\gamma(0)$~\cite{akutsu87}.

To calculate RHS of Eq.(\ref{tildef}), we treat the system by the transfer-matrix method.  Since the decorated RSOS model reduces to the original RSOS model, we can utilize the representation as the 19-state vertex model\cite{RSOS,honda} to make the transfer-matrix.  We can also use the decorated RSOS model itself by mapping the model to a ``decorated'' vertex model (Fig. \ref{vertex})\cite{akutsu99,akutsu00} with $19\times 16=304$ non-zero vertex weights (304-vertex model).  For approximate diagonalization of the transfer matrix, we employ the PWFRG method\cite{pwfrg,HOA,OHA,akutsu98} which is a variant of White's density matrix renormalization group (DMRG) method\cite{white}. 

From $\tilde{f}(\vec{\eta})$, we can calculate $f(\vec{p})$ (the surface free energy with fixed surface gradient vector $\vec{p}$), through the Legendre transformation:
\beq
f(\vec{p})=\tilde{f}(\vec{\eta}) + \vec{p} \vec{\eta} .\label{fpdef}
\eeq
We should recall the following thermodynamical equations associated with the Legendre transformation\cite{andreev}:
\beqa
p_x = - \partial \tilde{f}(\vec{\eta}) / \partial \eta_x,&& \  p_y = - \partial \tilde{f}(\vec{\eta}) / \partial \eta_y, \label{pdef}\\ 
\eta_x = \partial f(\vec{p}) / \partial p_x,&& \  \eta_y = \partial f(\vec{p}) / \partial p_y   \label{etadef}. 
\eeqa

In the actual calculation, we obtain $p = p_x(\eta)$ as the statistical average
\beq
p(\eta) = <\Delta h_x(m,n)> = <h{(m+1,n)}-h{(m,n)}>.
\eeq
By sweeping the field $\eta$, we obtain a $p-\eta$ curve.  In order to obtain $\gamma$ and $\tilde{\gamma}$, we use the relations (\ref{GMPTform}) and (\ref{univrel}).  Substituting Eq. (\ref{GMPTform}) into Eq. (\ref{etadef}), we have (for $\theta = 0$)
\begin{equation}
\eta = \gamma(0) +3B(0) |p|^2+ (\mbox{higher order}). \label{eta-p}
\end{equation}
Hence, we obtain $\gamma(0)$ and $B(0)$ by performing the least-square fitting of the $\eta-p$ curve (inverted $p-\eta$ curve).  The step stiffness $\tilde{\gamma}(0)$ is calculated from $B(0)$ through Eq. (\ref{univrel}).

\subsection{Surface free energies and surface gradient}

In Fig. \ref{bfp_fig} (b), we show $\tilde{f}(\eta)$ ($\eta_x = \eta, \eta_y=0$)  at several temperatures for $\alpha=0.5$ and $H=0$. The $p-\eta$ curves are shown in Fig. \ref{p-eta_fig}.  The roughening temperature is determined by Eq. (\ref{j0TR}) as $\kB\TR/\epsilon=1.446$ (\S IV).  At temperatures lower than $\TR$, the square-root behavior $p \sim (\eta-\gamma)^{1/2} $ characteristic to the GMPT universality class is seen.

From Eq. (\ref{fpdef}), we obtain $f(p)$ which is shown in Fig. \ref{bfp_fig} (a).  At $T<\TR$, $f(p)-p$ curve has a cusp at $p=0$.

\subsection{Adsorbate concentration}

The magnetization $M$ is calculated as $M= (M_x + M_y)/2$ where $M_x = <\sigma_x>$ and  $M_y = <\sigma_y>$ are directly calculated by the PWFRG method.  The adsorption coverage $\Theta$ is given by $\Theta=(M+1)/2$.  We have the following expressions for $M_x$ and $M_y$:
\beqa
M_x &=& <\sigma_x> = \frac{\sum_{\{ h{(i,j)} \} }\tanh \beta H_x^{\rm eff}(i,j) \exp[-\beta {\cal H}^{\rm eff}]}{\sum_{\{ h{(i,j)} \} }\exp[-\beta {\cal H}^{\rm eff}]},  \nonumber \\ 
&& \equiv <\tanh \beta H_x^{\rm eff}(i,j)>_{h},  \nonumber \\ 
M_y &=& <\sigma_y>  = <\tanh \beta H_y^{\rm eff}(i,j)>_{h}.    
\eeqa
Here
\beqa
H_x^{\rm eff}(i,j) &=& H + \epsilon \alpha |\Delta h_y{(i,j)}|,  \nonumber \\ 
&&\Delta h_y{(i,j)}=h(i,j+1)-h(i,j), \nonumber \\
H_y^{\rm eff}(i,j) &=& H + \epsilon \alpha |\Delta h_x{(i,j)}|,  \nonumber \\
 &&\Delta h_x{(i,j)}=h(i+1,j)-h(i,j), 
\eeqa
can be interpreted as the  effective local (magnetic) fields modulated by surface undulation.

Since $\Delta h$ takes $0$ or $\pm 1$, we have the identity $<F(|\Delta h|)>=F(0)+(F(1)-F(0))<|\Delta h|>$, which allows us to express $M_x$ and $M_y$ as
\beqa
M_x &=& \tanh \beta H + [\tanh \beta (H+\epsilon \alpha ) \nonumber \\
&& -\tanh \beta H]<|\Delta h_y{(i,j)}|>_{h},  \nonumber \\
M_y &=& \tanh \beta H + [\tanh \beta (H+\epsilon \alpha ) \nonumber \\
&& -\tanh \beta H]<|\Delta h_x{(i,j)}|>_{h}. \label{sigma_av}
\eeqa
We should note the relations,
\beqa
<|\Delta h_y{(i,j)}|>_{h} &=& <\delta_{\Delta h_y{(i,j)},1}>_{h}  \nonumber \\
&&+<\delta_{\Delta h_y{(i,j)},-1}>_{h}, \nonumber \\
<|\Delta h_x{(i,j)}|>_{h} &=& <\delta_{\Delta h_x{(i,j)},1}>_{h}  \nonumber \\
&&+<\delta_{\Delta h_x{(i,j)},-1}>_{h},\label{Delta_h_av}
\eeqa
where $\delta_{i,j}$ is the Kronecker's delta.
Then, we have
\beqa
p_y &=& <\delta_{\Delta h_y{(i,j)},1}>_{h}-<\delta_{\Delta h_y{(i,j)},-1}>_{h}=0,\nonumber \\
p_x &=& <\delta_{\Delta h_x{(i,j)},1}>_{h}-<\delta_{\Delta h_x{(i,j)},-1}>_{h}=p.\label{pxpy}
\eeqa
Substituting Eq. (\ref{Delta_h_av}) and Eq. (\ref{pxpy}) into Eq. (\ref{sigma_av}), we obtain
\beqa
M_x &=& \tanh \beta H + [\tanh \beta (H+\epsilon \alpha ) \nonumber \\
&& -\tanh \beta H] \cdot 2<\delta_{\Delta h_y{(i,j)},-1}>_{h}, \nonumber \\
M_y &=& \tanh \beta H + [\tanh \beta (H+\epsilon \alpha ) \nonumber \\
&& -\tanh \beta H](p+2<\delta_{\Delta h_x{(i,j)},-1}>_{h}). \label{Meq}
\eeqa

In Fig.~\ref{m-p_fig}, we show $p$-dependence of $M_x$ and $M_y$ for $\alpha=0.5$ and $H=0$.  As is seen from the figure, $<\delta_{\Delta h_y{(i,j)},-1}>_{h}$ depends on $p$, taking maximum around $p \sim 0.5$.  At $p=0$, $<~\delta_{\Delta h_y{(i,j)},-1}>_{h}$ ($= <\delta_{\Delta h_x{(i,j)},-1}>_{h}$) is small because the surface is in the smooth phase; while at $p=1$, $<~\delta_{\Delta h_y{(i,j)},-1}>_{h}=0$ due to the geometrical restriction in the RSOS model.  Therefore,  $<\delta_{\Delta h_y{(i,j)},-1}>_{h}$ should have a  maximum in between  $p=0$ and $p=1$.

\subsection{Step tension and step stiffness}

At low temperatures, step quantities of the RSOS model are well approximated by the interface quantities of 2D Ising model\cite{akutsu98}.
  Using the Ising-model results, we have the following approximate analytic expressions of  $\gamma$ and $\tilde{\gamma}$\cite{rottman81,akutsu86}, 
\beqa
\gamma&=& \epsilon^{\rm eff}-\kBT \ln [\tanh (\epsilon^{\rm eff}/2\kBT)],\label{gameq} \\
\tilde{\gamma}&=&\kBT \sinh (\gamma/\kBT) \nonumber \\
&=& \kBT[\exp(\epsilon^{\rm eff}/\kBT)\tanh(\epsilon^{\rm eff}/(2\kBT)) \nonumber \\
&&-\exp(-\epsilon^{\rm eff}/\kBT)\coth(\epsilon^{\rm eff}/(2\kBT))]/2.\label{stiffeq}
\eeqa

In Fig. \ref{gam_fig} and \ref{stiff_fig}, we show the  temperature  dependence of step tension and step stiffness given by Eq. (\ref{gameq}) and (\ref{stiffeq}) for several values of $H$ (or, equivalently, $\mu/2$).  We see that the Ising-model expressions (\ref{gameq}) and (\ref{stiffeq}) agree very well with the PWFRG calculation at low temperatures.

In the temperature dependence of $\gamma$, there is an interesting property that $\gamma$ exhibits the quasi-reentrant behavior; there appears a temperature region where $\gamma $ increases as we raise the temperature, although $\gamma$ is usually a decreasing function of the temperature.  Accordingly, $\gamma$ has a maximum at a temperature above which $\gamma$ decreases again as we raise the temperature.  This behavior of $\gamma$ is what we call the quasi-reentrant behavior of a step, which can be seen in the 2D frustrated Ising models\cite{akutsu96}. Since the size of the facet in the ECS is proportional to the step tension, the facet size should also exhibit the quasi-reentrant behavior\cite{akutsu96}.

The step tension and step stiffness in the lattice-gas representation are given in Fig. \ref{gamlg_fig} and \ref{stifflg_fig}, respectively.  When $\alpha=0.5$ (Fig. \ref{gamlg_fig} (b) and Fig. \ref{stifflg_fig} (b)), $\gamma$ and $\tilde{\gamma}$ of the surface with adsorption are smaller than those of the clean surface;  the step fluctuation width which is inversely proportional to $\tilde{\gamma}$~\cite{akutsu86,sig-gam} becomes larger, and $\TR$ becomes lower.  As the vapor pressure $P$ is raised, $\gamma$ and $\tilde{\gamma}$ become smaller because the increase of adsorbase concentration causes the decrease of $\epsilon^{\rm eff}$ (Fig. \ref{eefflg_fig} (b)).  At $\alpha=-0.5$, $\gamma$ and $\tilde{\gamma}$ of adsorbed surface are larger than those of clean surface, and $\TR$  becomes higher.  As we raise $P$, $\gamma$ and $\tilde{\gamma}$  grows, because the increase of adsorbate concentration causes the increase of $\epsilon^{\rm eff}$ (Fig. \ref{eefflg_fig} (e)).

\section{Summary}

In this paper, we have discussed the Langmuir adsorption effect on the vicinal surface below the roughening temperature, in terms of the restricted solid-on-solid (RSOS) model coupled with the Ising spins.  We have shown that this ``decorated'' RSOS model reduces to the ordinary RSOS model with the effective ledge energy.  The effective ledge energy depends on temperature $T$ and the chemical potential $\mu$ of the adsorbates, which is caused by the change of the adsorbate concentration.

We have found that, in a parameter region, there occurs novel multi-reentrant roughening transition.  By employing the product-wave-function renormalization group (PWFRG) method, we have calculated thermodynamical quantities associated with the vicinal surface of the model.  We have also obtained quasi-reentrant behavior in the temperature dependence of step tension in a wide area of the phase diagram.

We should give a comment on a related study made previously. In Ref. \cite{kariotis}, a similar modeling is adopted for the absolute SOS model and the discrete Gaussian model, where the adsorption effect on the roughening temperature is discussed in terms of the approximate renormalization group approach (vicinal surface problem is not discussed).  One of the merits of our RSOS modeling is that exact renormalization of coupling constant is available, which allows us semi-analytical treatment of the problem.

\acknowledgements

 This work was partially supported by the ``Research for the Future'' Program from The Japan Society for the Promotion of Science (JSPS-RFTF97P00201) and by the Grant-in-Aid for Scientific Research from Ministry of Education, Science, Sports and Culture (No.12640393).

\begin{figure}[htbp]
\caption{ (a) RSOS heights ($h_{1},\ldots, h_{4}$) and the edge variables in the mapped vertex model. (b) A vertex and leg variables.  (c) A vertex decorated by spins which are represented by circles.}
\label{vertex}
\end{figure}

\begin{figure}[htbp]
\caption{
Temperature dependence of the effective coupling constant $\epsilon^{\rm eff}$ in the spin representation (Eq. (3.3)). (a) $\alpha=0.1$. (b) $\alpha=0.5$. (c) $\alpha=1.$
In each figure, we draw the line for $H/\epsilon=0.2$ (dashed), $H/\epsilon=0.1$ (dotted), $H/\epsilon=0$ (solid), $H/\epsilon=-0.1$ (two-dot dashed), and $H/\epsilon=-0.2$ (dot dashed).
} \label{eeff_fig}
\end{figure}

\begin{figure}[htbp]
\caption{
Temperature dependence of the effective ledge energy $\epsilon^{\rm eff}$ in the lattice gas representation (Eq. (3.5)). 
In each figure, we draw the line for $H/\tilde{\epsilon}=0.2$ (dashed), $H/\tilde{\epsilon}=0.1$ (dotted), $H/\tilde{\epsilon}=0$ (solid), $H/\tilde{\epsilon}=-0.1$ (two-dot dashed), and $H/\tilde{\epsilon}=-0.2$ (dot dashed).
(a) $\tilde{\alpha}=0.2$. (b) $\tilde{\alpha}=0.5$. (c) $\tilde{\alpha}=1$. (d) $\tilde{\alpha}=-0.2$.  (e) $\tilde{\alpha}=-0.5$. (f) $\tilde{\alpha}=-1$.
} \label{eefflg_fig}
\end{figure}

\begin{figure}[htbp]
\caption{
Determination of the roughening transition temperature $\TR$ (Eq. 4.1).
(a) $\alpha=0.5$. $H/\epsilon=0.2$ (dashed), $H/\epsilon=0.1$ (dotted), $H/\epsilon=0$ (thick solid), $H/\epsilon=-0.1$ (two-dot dashed), and $H/\epsilon=-0.2$ (dot dashed).
(b) $\alpha=1.0$. $H/\epsilon=0.02$ (dashed), $H/\epsilon=0.01$ (dotted), $H/\epsilon=0$ (thick solid), $H/\epsilon=-0.01$ (two-dot dashed), and $H/\epsilon=-0.02$ (dot dashed).
(c) $\alpha=1.1$
 $H/\epsilon=-0.0685$ (dashed), $H/\epsilon=-0.0735$ (dotted), and $H/\epsilon=-0.0785$ (dot dashed).
The horizontal line (thin solid) corresponds to  $\zeta_{\rm R} = \epsilon^{\rm eff}/\kBT=0.633$.
} \label{j0TR_fig}
\end{figure}

\begin{figure}[htbp]
\caption{
Phase diagram.
(a) $\alpha=0.5$.
(b) $\alpha=1.0$.
(c) $\alpha=1.1$.
(d) $\alpha=1.2$.
} \label{phaseTR_fig}
\end{figure}

\begin{figure}[htbp]
\caption{
(a) Temperature dependence of the  $f(p)/\kBT$. 
(b) Temperature dependence of the Andreev free energy $\tilde{f}(\eta)$ for $\alpha=0.5$ and  $H= 0$. From top to bottom,  $\kBT/\epsilon = 0.4$, 0.6, 0.8, 1.0 and 1.2.
}
\label{bfp_fig}
\end{figure}

\begin{figure}[htbp]
\caption{
PWFRG results for the $p$-$\eta $ curves ($\alpha=0.5$, and $H=0$).   Temperature of each curve is $\kBT/\epsilon =$ 0.4, 0.6, 0.8, 1, 1.2, and 1.4, from right to left.
} \label{p-eta_fig}
\end{figure}

\begin{figure}[htbp]
\caption{
$p$-dependence of $M_x$ (solid) and $M_y$ (dashed) for  $\alpha=0.5$ and $H= 0$. From top to bottom at $p=0$,  $\kBT/\epsilon = 1.1$, 0.8 and 0.5.
}
\label{m-p_fig}
\end{figure}

\begin{figure}[htbp]
\caption{
Temperature-dependence of the step tension $\gamma$ (Eq. (5.16)).
Open circles correspond to PWFRG calculation for  $\alpha=0.5$ and $H=0$.
We draw the line for $H/\epsilon=0.2$ (dashed), $H/\epsilon=0.1$ (dotted), $H/\epsilon=0$ (solid), $H/\epsilon=-0.1$ (two-dot dashed), and $H/\epsilon=-0.2$ (dot dashed).
(a) $\alpha=0.1$. (b) $\alpha = 0.5$.
} \label{gam_fig}
\end{figure}

\begin{figure}[htbp]
\caption{
Temperature-dependence of the step stiffness $\tilde{\gamma}$ (Eq. (5.17)).
Open circles correspond to PWFRG calculation for  $\alpha=0.5$ and $H=0$.
We draw the line for $H/\epsilon=0.2$ (dashed), $H/\epsilon=0.1$ (dotted), $H/\epsilon=0$ (solid), $H/\epsilon=-0.1$ (two-dot dashed), and $H/\epsilon=-0.2$ (dot dashed).
(a) $\alpha=0.1$. (b) $\alpha=0.5$.
} \label{stiff_fig}
\end{figure}

\begin{figure}[htbp]
\caption{
Temperature-dependence of the step tension $\gamma$ (Eq. (5.16) and (3.5)) in the lattice gas representation. 
We draw the line for $H/\tilde{\epsilon}= 0.2$ (dashed), $H/\tilde{\epsilon}= 0.1$ (dotted), $H/\tilde{\epsilon}= 0$ (solid), $H/\tilde{\epsilon}= -0.1$ (two-dot dashed), and $H/\tilde{\epsilon}= -0.2$ (dot dashed).
(a) The upper group of the line corresponds to $\tilde{\alpha}= -0.2$ and the lower group  to  $\tilde{\alpha}= 0.2$.
 (b) The upper group of the line corresponds to  $\tilde{\alpha}= -0.5$ and the lower group to $\tilde{\alpha}= 0.5$.
} 
\label{gamlg_fig}
\end{figure}

\begin{figure}[htbp]
\caption{
Temperature-dependence of the step stiffness $\tilde{\gamma}$ (Eq. (5.17) and (3.5)) in the lattice gas representation. 
We draw the line for $H/\tilde{\epsilon}= 0.2$ (dashed), $H/\tilde{\epsilon}= 0.1$ (dotted), $H/\tilde{\epsilon}= 0$ (solid), $H/\tilde{\epsilon}= -0.1$ (two-dot dashed), and $H/\tilde{\epsilon}= -0.2$ (dot dashed).
(a) The upper group of the line corresponds to  $\tilde{\alpha}= -0.2$ and the lower group to $\tilde{\alpha}= 0.2$.
 (b) The upper group of the line corresponds to $\tilde{\alpha}= -0.5$ and the lower group to $\tilde{\alpha} = 0.5$.
} \label{stifflg_fig}
\end{figure}


\begin{references}
\bibitem{Theta}
  If we set up the $xy$ coordinates on the facet plane so that the $y$-axis corresponds to a crystal axis, surface direction of a vicinal surface is specified by the surface gradient vector $\vec{p} = (p_x, p_y)$ where $p_{x}$ (resp. $p_{y}$) is the mean surface gradient along the $x$- (resp. $y$-) direction.  The step density $\rho$ relates to $\vec{p}$ by $\rho = |\vec{p}|/a_h = \sqrt{{p_x}^2 + {p_y}^2} /a_h$ where $a_{h}$ is the height of a single step.  The mean running direction angle $\theta$ of the steps relates to $\vec{p}$ as $\vec{p}=|\vec{p}|(\cos\theta, \sin\theta)$. We regard the free energy $f$ as a function of the gradient vector $\vec{p}=(p_x,p_y)$, allowing us to write $f = f(\vec{p})$. In the present paper, we take the units where $a_{h}=1$.  In the actual calculation, we consider the case where the steps are running along the $y$-direction ($p_y=0$).
\end{references}
\end{document}